\documentclass[sigconf]{acmart}

\copyrightyear{2026}
\acmYear{2026}
\setcopyright{cc}
\setcctype{by}
\acmConference[ICAIF '26]{7th ACM International Conference on AI in Finance}{November 14--17, 2026}{Milan, AA, Italy}
\acmBooktitle{7th ACM International Conference on AI in Finance (ICAIF '26), November 14--17, 2026, Milan, AA, Italy}
\acmDOI{10.1145/3858063.3858865}
\acmISBN{979-8-4007-3087-0/2026/11}

\makeatletter\@ifundefined{KV@Gin@artifact}{\define@key{Gin}{artifact}[]{}}{}\makeatother
\graphicspath{{./}}
\setkeys{Gin}{width=\linewidth,keepaspectratio}

\begin{document}

\title{Algorithmic Collusion Without Deterrence: Auditing the Punishment Mechanism in Q-Learning Pricing Agents}

\author{Kau\^{e} Lopes de Moraes}
\orcid{0000-0002-2904-7625}
\affiliation{
  \institution{University of S\~{a}o Paulo}
  \city{S\~{a}o Paulo}
  \country{Brazil}}
\email{kaue.moraes@usp.br}

\author{Rodrigo De Losso}
\orcid{0000-0002-3461-8621}
\affiliation{
  \institution{University of S\~{a}o Paulo}
  \city{S\~{a}o Paulo}
  \country{Brazil}}
\email{delosso@usp.br}

\renewcommand{\shortauthors}{Lopes de Moraes and De Losso}
\begin{abstract}
Independent tabular Q-learning agents repricing in a simulated Bertrand
duopoly converge to supracompetitive prices, and the seminal evidence
attributes this to a learned reward--punishment scheme: undercut your
rival and it retaliates, so deviation never pays. This mechanistic
reading is what turns high prices into a cartel, and into a regulatory
target. We replicate the environment of \citet{calvano2020} and confirm
their impulse-response pattern within co-trained pairs. We then audit
the pattern on frozen policies, with learning off, and find that on its
own it does not identify a mechanism: the return is over-determined
rather than selective. In nine of ten converged pairs (914 of 1,024) a
single attractor's basin is the entire state space, so every
perturbation returns whatever provoked it, and a forced price
\emph{increase} triggers the same retaliatory pattern as an undercut.
What co-training produces is a narrow, dyad-specific basin of
attraction: re-match frozen agents from independent runs, as Calvano et
al.~did with learning on, and coordination collapses by 80\% with
nothing carried across. The deterrence claim is incomplete on its own
terms: the one-period undercut of Calvano et al.'s impulse-response test
is unprofitable, which we confirm in 92\% of pairs, but the
\emph{optimal} deviation, computed by dynamic programming against each
partner's frozen strategy, profits by 17.5\% on average. The scheme
deters only the deviation no optimizer would choose; collusion survives
because deviation is never tried, not because it would not pay.
Supracompetitive prices are robust; the punishment mechanism read into
them does not survive these tests, which changes what the standard test
of algorithmic collusion establishes, and what pairing-disruption
remedies can be expected to do.
\end{abstract}



\maketitle

\section{Introduction}\label{introduction}

A growing literature reports that simple reinforcement-learning agents,
trained independently in repeated pricing games, converge to prices well
above the static Nash equilibrium. The seminal contribution of
\citet{calvano2020} goes further than documenting supracompetitive
prices: it offers a mechanism. After a forced unilateral price cut, the
rival's price drops and both prices then return to their pre-deviation
level, a pattern the authors read as a ``reward--punishment scheme''
that stabilizes the collusive outcome. This mechanistic reading is what
gives the result its policy weight: prices that are merely high are an
artifact; prices \emph{defended by retaliation} are a cartel.

This paper asks a narrow question with broad consequences: \textbf{is
there actually a punishment mechanism?} We take the question seriously
on Calvano et al.'s own terms. We replicate their environment and their
impulse-response experiment, and we confirm their pattern: within
co-trained pairs, the deviation-drop-return dynamic is robust (Section
4.2). The contribution builds on a line of work that has questioned
parts of the mechanistic reading with separate instruments (Section 2),
and it is twofold. The negative half is a five-part audit of the
mechanism behind the pattern (Section 4.3) and an incentive accounting
of the converged play (Section 4.5), run jointly on the same frozen
policies: taken on its own terms, the exhibit does not establish
deterrence. The positive half is what we find in its place: co-training
produces a \emph{dyad-specific basin of attraction}, a joint object
carved by the pair and without value outside it. The direct test
re-matches converged agents across independent training runs with
learning off, as \citet{eschenbaum2022} do in a 20-price variant of the
environment, and runs it at scale on the same frozen policies we audit:
coordination collapses by 80\% (Section 4.4). Throughout, the design
separates the \emph{learning phase} from the \emph{measurement phase}:
in the language of experimental economics, Calvano et al.~evaluate in
the partner treatment, and we add the stranger treatment as a test
rather than as a re-training exercise (Section 3.1).

None of the five audits finds a punishment mechanism. The retaliatory
response fires for \emph{any} off-path price, including a cooperative
price increase; faced with an opponent that actually punishes, the
agents undercut it and absorb the permanent punishment rather than
sustain the cooperative path it offers; the learned reaction functions
are only weakly and inconsistently related to best-response logic; the
celebrated return-to-baseline is forced by the geometry of the converged
policies, which in nine pairs of ten have a single global attractor, so
it would follow any perturbation; and none of this reflects a
representational limit, because hand-built punishment rules fit the same
policy matrix exactly. What survives the audit is real but different in
kind: a locally calibrated, dyad-specific basin of attraction, fast to
return, blind to why it was perturbed, and specific to the pair that
carved it. Deterrence is a property an agent has; what training built
here is a property only the pair has.

We are careful about what we do \emph{not} claim. Calvano et al.~report
that about half of their sessions reach an on-path Nash equilibrium of
the repeated game, with small one-step losses elsewhere, and rarely a
subgame-perfect one; we do not dispute these statistics, and Section 4.5
says how our best-response measure relates to theirs. Our target is the
mechanistic interpretation, the ``reward--punishment scheme,'' in the
sense that matters for economics: a scheme, in the tradition of
\citet{friedman1971} and \citet{abreu1988}, is a rule that
\emph{enforces a prescribed path}. It responds to departures from that
path, and it sustains the path because the response makes deviating
unprofitable. That enforcement structure is what the audits do not find.

\section{Background}\label{background}

\paragraph{Punishment schemes.}\label{punishment-schemes.}

The language of ``reward--punishment schemes'' carries a precise meaning
in repeated-game theory. In \citet{friedman1971}, collusion is sustained
by a trigger strategy: cooperate until anyone defects, then revert to
static Nash forever. \citet{abreu1988} generalizes this to optimal penal
codes, in which deviations, including deviations from the punishment
itself, restart a calibrated punishment phase. Two features are
constitutive. First, \emph{path enforcement}: the scheme prescribes a
path and responds to departures from it, in order to make deviating
unprofitable. Second, \emph{incentive compatibility}: the threat is
credible because carrying it out is optimal given the scheme (under
imperfect monitoring, punishment can even fire without any deviation;
\citet{green1984}). Our audit therefore does not ask whether the learned
response fires only after defections (a trigger rule need not be
selective in that sense) but whether the response has the
\emph{structure} of path enforcement (Sections 4.3(i)--(iii)), whether
the enforced path is incentive-compatible (Section 4.5), and whether the
return itself is selective at all, or forced by the structure of the
state space (Section 4.3(iv)).

\paragraph{\texorpdfstring{The \citet{calvano2020}
result.}{The @calvano2020 result.}}\label{the-calvano2020-result.}

Two firms in a logit-Bertrand duopoly simultaneously choose prices from
a 15-point grid; demand is
\(q_i = e^{(a_i-p_i)/\mu} / \big(\sum_j e^{(a_j-p_j)/\mu} + e^{a_0/\mu}\big)\)
and profit is \(\pi_i = (p_i - c)\,q_i\), with \(a=2\), \(c=1\),
\(\mu=0.25\), \(a_0=0\). Each agent runs tabular Q-learning
\citep{watkins1992} on the state \(s_t = (p^A_{t-1}, p^B_{t-1})\) (225
states) with \(\alpha = 0.15\), \(\delta = 0.95\), and
\ensuremath{\varepsilon}-greedy exploration decaying geometrically to
zero. After convergence, prices settle well above Nash. We measure this
with the collusion index
\(\mathrm{CI} = (\bar{p}^{RL} - p^{NE})/(p^{M} - p^{NE})\), where
\(p^{NE} \approx 1.473\) and \(p^{M} \approx 1.925\). Calvano et al.'s
impulse-response analysis (force one agent to undercut for one period;
the rival's price falls and both then recover) is the exhibit read as a
``reward--punishment scheme,'' and it is what elevated the result to a
landmark for competition policy.

\paragraph{Subsequent literature and the
gap.}\label{subsequent-literature-and-the-gap.}

\citet{klein2021} finds supracompetitive outcomes under sequential
pricing, sensitive to action-space granularity. \citet{asker2024} show
that, for myopic agents, counterfactual (``synchronous'') updating
drives prices to the competitive level, so that high prices stem from
what the learning protocol fails to evaluate rather than from strategy;
\citet{hansen2021} obtain supracompetitive prices from bandit algorithms
with \emph{no} punishment capability, and \citet{abada2023} reach a
similar exploration-based diagnosis. Empirically, algorithmic pricing
changes price competition \citep{brown2023, assad2024}; the regulatory
stakes are drawn out by \citet{harrington2018} and \citet{oecd2017}.
Closest to us, \citet[hereafter Calvano IJIO]{calvano2023} classify the
collusion learned under the 2020 exploration protocol as \emph{genuine},
on the strength of its impulse responses and its dependence on the
discount factor, against the \emph{spurious} high prices of optimistic
initialization. The interpretation has since been examined from several
directions, each with its own instrument. \citet{epivent2024} show, in
the same environment, that the price war follows price \emph{hikes} as
well as cuts, so the response does not distinguish the direction of the
deviation. \citet{denboer2022} show that the impulse-response pattern
(the rival cuts, prices return, the deviation does not pay) can be
generated by strategy profiles that are neither equilibria nor
reward--punishment schemes, and conclude that a systematic test would
have to check every unilateral deviation; they carry this out
exhaustively for two prices, and the 15-price case is beyond
enumeration. \citet{banchio2022} describe a mechanism that sustains
supracompetitive prices without any punishment: correlated estimation
errors across co-learning agents, which they call spontaneous coupling.
And \citet[Section VI.B]{calvano2020} themselves rematch converged
agents with learning still on: profit gains fall from 85\% to about 20\%
before re-learning partially restores them, which they read as
coordination being ``almost completely pair-specific.''
\citet{eschenbaum2022} run the rematch with learning off, in the same
environment on a 20-price grid: policies from sessions that differ only
in the seed revert to Nash play on average, which they attribute to
overfitting to the training opponent, and restricting each agent's state
to its own last price makes collusion robust across contexts.
\citet{tang2024} surveys this debate under the labels \emph{spurious},
\emph{genuine} and \emph{authentic} collusion, the last being Tang's
name for the definition of \citet{denboer2023}.

What has been missing is a joint audit, on the same converged policies
and with learning off, that applies these instruments together and adds
the three the debate still lacks: an explicit random-policy null for the
return itself, a capacity control for whether the architecture could
express a punishment scheme at all, and an exact best response against
each converged partner, the equilibrium check den Boer et al.~perform by
enumeration for two prices, carried out in the 15-price case. The
cross-pairing control follows \citet{eschenbaum2022} in using rematching
with learning off as an evaluation protocol; we run it at scale and on
the same policies as the mechanism audit, and connect it to the
brittleness documented in multi-agent RL as joint policy correlation,
the overfitting of independently trained policies to their co-learners
\citep{lanctot2017}.

\section{Experimental Design}\label{experimental-design}

\subsection{Training and testing: the logic of the
design}\label{training-and-testing-the-logic-of-the-design}

Every result in this paper rests on the separation between a
\emph{learning phase} and a \emph{measurement phase}. The training stage
(learning ON) is not the object of study; it is the process that
generates the objects of study, the analogue of the early rounds of a
laboratory experiment, or of the estimation sample in econometrics. At
convergence, the Q-tables are frozen and play becomes greedy: the test
stage (learning OFF), the analogue of out-of-sample validation. Every
number in this paper is collected in the test stage.

This separation is where the literature's evidence is thinnest. In the
standard framework (Figure 1), the pair that learns together is the same
pair that is evaluated after convergence: all impulse-response evidence,
in \citet{calvano2020} and in the studies that followed it, is collected
\emph{within the original training dyad}, the \emph{partner design} of
experimental economics \citep{andreoni1988}. Nothing in that design can
distinguish an individual strategy, which an agent carries with it, from
a joint convention, which exists only inside the pair. In
machine-learning terms, the standard test evaluates each policy exactly
on the distribution of states its own training partner generates: an
in-sample measurement, presented as evidence about the policy itself.
But a mechanism claim is a claim about the \emph{agent}: that it carries
a rule that detects and punishes deviations. Testing that claim requires
moving the agent out of the context that produced it. \citet[Section
VI.B]{calvano2020} do move it, as a re-training exercise with learning
on; \citet{eschenbaum2022} use the move as a test with learning off, so
that what the policy carries is separated from what re-learning
recovers.

\begin{figure}
\centering
\includegraphics[keepaspectratio]{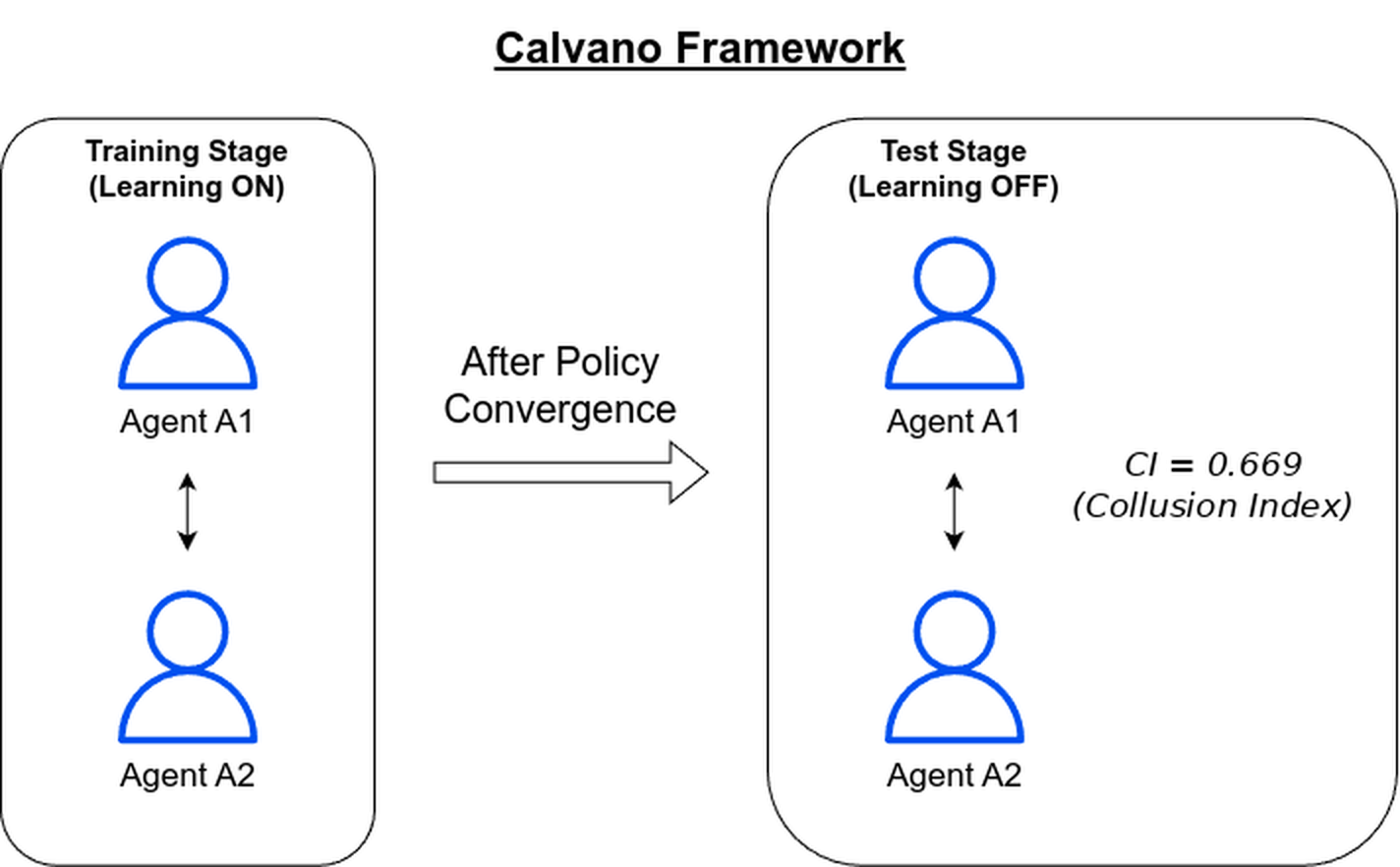}
\Description{Diagram of the standard evaluation framework. Two agents, A1 and A2, learn together in a training stage with learning on, and after policy convergence the same two agents are evaluated in a test stage with learning off, where the collusion index is 0.669.}
\caption{The standard evaluation framework of the algorithmic collusion
literature \citep{calvano2020}: the pair of agents that learns together
(training stage, learning ON) is the same pair that is evaluated after
policy convergence (test stage, learning OFF): the partner design. All
impulse-response evidence is collected within the original training
dyad; training and testing contexts are never separated. The collusion
index shown (CI = 0.669) is the within-pair value from our replication.}
\end{figure}

We adopt that manipulation as a test: a one-shot \emph{stranger}
evaluation of frozen converged play (Figure 2). After convergence, an
agent from run A is paired with an agent from run B, a partner it never
interacted with. The treatment is the pairing, not the training, and the
question it answers is the classic rematching question: is the observed
behavior an individual strategy, or a convention of the pair? Freezing
is not a simplification but the regime that matters for deployment: the
policy that reaches a market is the one trained offline, and Calvano et
al.'s rematched agents re-coordinate only after tens of thousands of
further periods of greedy play at live prices against the new competitor
(their Figure 11; fast relative to the original training, as they note,
but paid in realized profit). The frozen test asks what the policy
carries before that bill is paid.

\begin{figure}
\centering
\includegraphics[keepaspectratio]{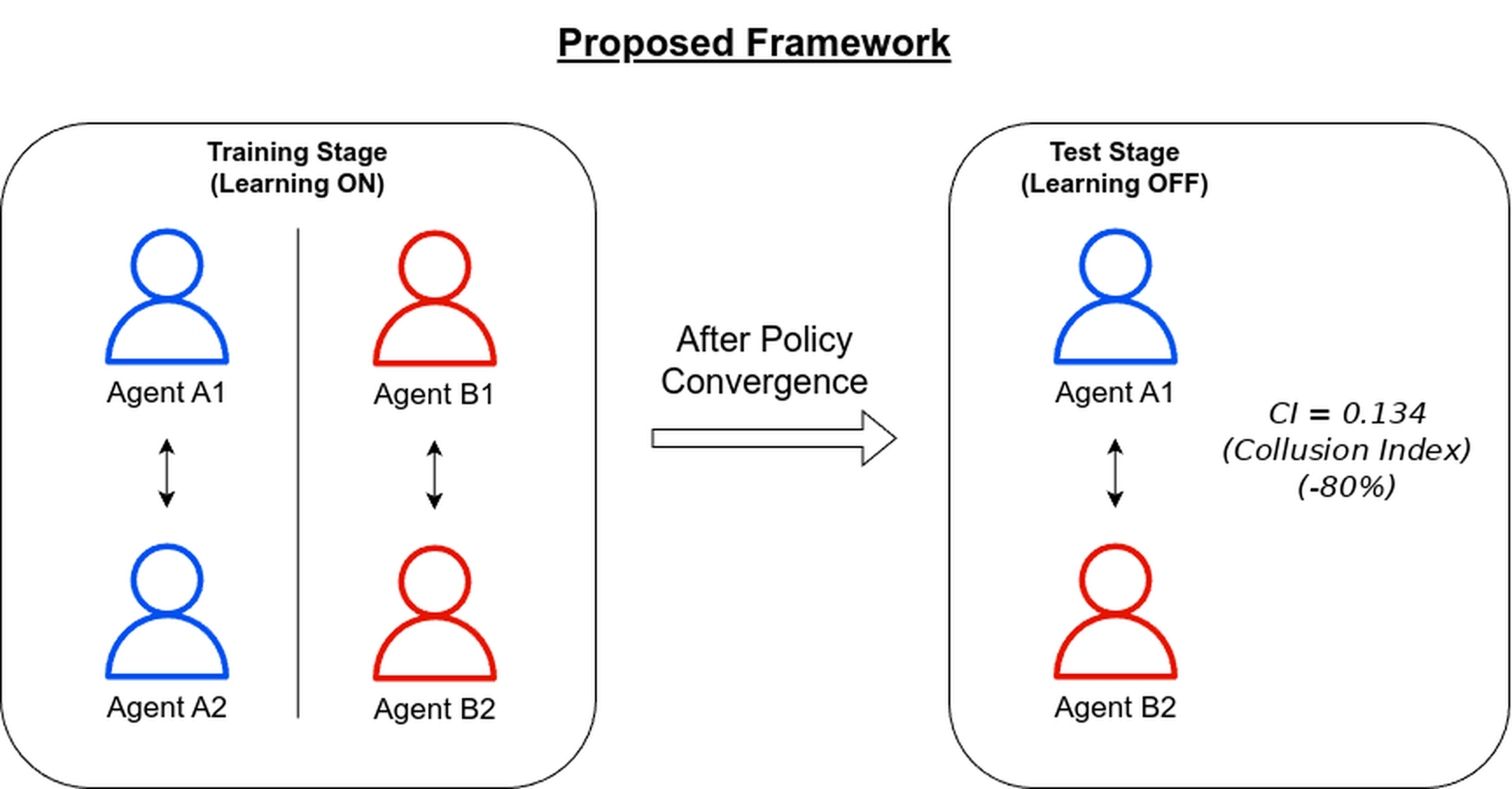}
\Description{Diagram of the stranger test. Two pairs, A1 with A2 and B1 with B2, train separately with learning on. After convergence the test stage pairs A1 with B2, agents that never interacted, with learning off; the collusion index falls to 0.134, 80 percent lower.}
\caption{The stranger test: Calvano et al.'s rematching confined to the
test stage, with learning off. Training is identical to Calvano et
al.'s, replicated in independent runs (run A, run B); the manipulation
is confined to the test stage: pairing an agent from run A with an agent
from run B it never interacted with. If co-training produced individual
collusive strategies, coordination would survive the re-pairing; if it
produces a dyad-specific joint basin, it collapses. It collapses: CI
falls from 0.669 (within-pair) to 0.134 (\ensuremath{-}80\%; Section
4.4).}
\end{figure}

\subsection{Environment, training, and
protocols}\label{environment-training-and-protocols}

\paragraph{Environment and training.}\label{environment-and-training.}

Bertrand-logit duopoly as in Section 2, on Calvano et al.'s 15-point
grid (prices from 1.43 to 1.97, the Nash--monopoly interval extended by
10\% on each side). Agents are trained in pairs (``co-trained'') for 2M
steps, in eight batches of 128 runs under an identical protocol with
distinct seeds. Each run produces one co-trained pair, and the core
population is the resulting 1,024 pairs (2,048 agents); every count in
Section 4 is over this population unless stated otherwise. All
post-training evaluations are greedy (no learning, no exploration).
Deliberate divergences from Calvano et al.'s protocol (a fixed 2M-step
budget with 10,000-step episode resets and random initial states, versus
their infinite horizon and 100,000-period stability criterion) are
controlled for by a separate five-pair population trained under the
faithful regime (no resets, single horizon, stability stopping rule), on
which Section 5 reruns the entire audit chain. In total the paper draws
on 1,024 co-trained pairs at 2M steps, one 10M-step reference run, five
pairs under the faithful regime, and 1,024 further pairs per grid size
in the granularity sweep of Section 4.6; every audit is computed exactly
on the converged policies, pair by pair, rather than averaged over
sessions.

\paragraph{Reproducibility.}\label{reproducibility.}

The environment, agents, training and audit scripts, configurations,
seeds and the converged Q-tables behind every figure are available at
\url{https://github.com/kauelopes/algorithmic-collusion-without-deterrence}.

\paragraph{Protocols.}\label{protocols.}

(i) \emph{Impulse response}: after the pair settles, one agent's price is
  exogenously forced for one period and both prices are recorded over a
  {[}\ensuremath{-}5, +40{]} window; shocks go \emph{down} (to the
  static best response, Calvano et al.'s deviation) and \emph{up} (to
  the monopoly price); each cell runs under two arms, co-trained and a
  \emph{cross-paired} control (random matchings across runs), with 10
  random initializations per pair. (ii) \emph{Fixed-opponent
  evaluation}: each trained agent plays nine scripted strategies; a
  dedicated opponent-model test initializes the joint state at the
  cooperative price, so that trigger strategies start untriggered, and
  plays each agent 50 periods against grim-trigger, constant-Nash and
  constant-monopoly opponents. (iii) \emph{Cross-deployment}: agents
  from different runs are matched and evaluated greedily. (iv)
  \emph{Exact dynamic programming}: fixing the partner's greedy policy
  induces a 225-state MDP; we compute the exact value of converged play
  and of the optimal best response; the partner \emph{reacts}, so this
  is not a one-shot counterfactual. (v) \emph{Null model}: the
  impulse-response protocol applied to 1,024 pairs of random Q-tables
  drawn on the same value range, five initializations each, the same
  trial count as the trained arm. (vi) \emph{Capacity control}:
  hand-constructed punishment policies embedded in the same 225
  \ensuremath{\times} 15 matrix.

\section{Results}\label{results}

\subsection{The phenomenon replicates}\label{the-phenomenon-replicates}

Everything we go on to audit, we first reproduce. Across the 1,024
co-trained pairs at 2M steps, the mean long-run collusion index is
\textbf{CI = 0.668} (s.d. 0.184; CI = 0 at the Nash price, 1 at the
monopoly price), within the range spanned by published replications; a
single high-fidelity run at 10M steps on the same grid yields CI =
0.6421. Our environment reproduces the phenomenon we set out to examine.
One number recorded here becomes important later: in 1,023 of the 1,024
pairs at least one agent has a strictly positive exact best-response
gain against its own partner (Section 4.5). With a single exception, no
converged profile is free of an incentive to deviate.

\subsection{The punishment pattern reproduces within the
pair}\label{the-punishment-pattern-reproduces-within-the-pair}

We first run Calvano et al.'s own test. Within co-trained pairs, a
forced one-period undercut produces exactly the published pattern: the
rival's price falls (in 1,013 of 1,024 pairs; mean drop 0.303 from a
pre-shock level of 1.775), and in 1,006 of 1,024 pairs (98.2\%) both
prices return to the pre-deviation attractor, exactly to the pre-shock
prices when the attractor is a fixed point (60\% of pairs) and to the
same cycle otherwise; 618 pairs are back within five periods, and the
mean is 4.7 (Figure 3, top-left). The impulse-response evidence in
Calvano et al. is not an artifact of their code or ours; it is a robust
property of co-trained Q-learning pairs in this environment. The
question is what it is evidence \emph{of}.

\begin{figure}
\centering
\includegraphics[keepaspectratio]{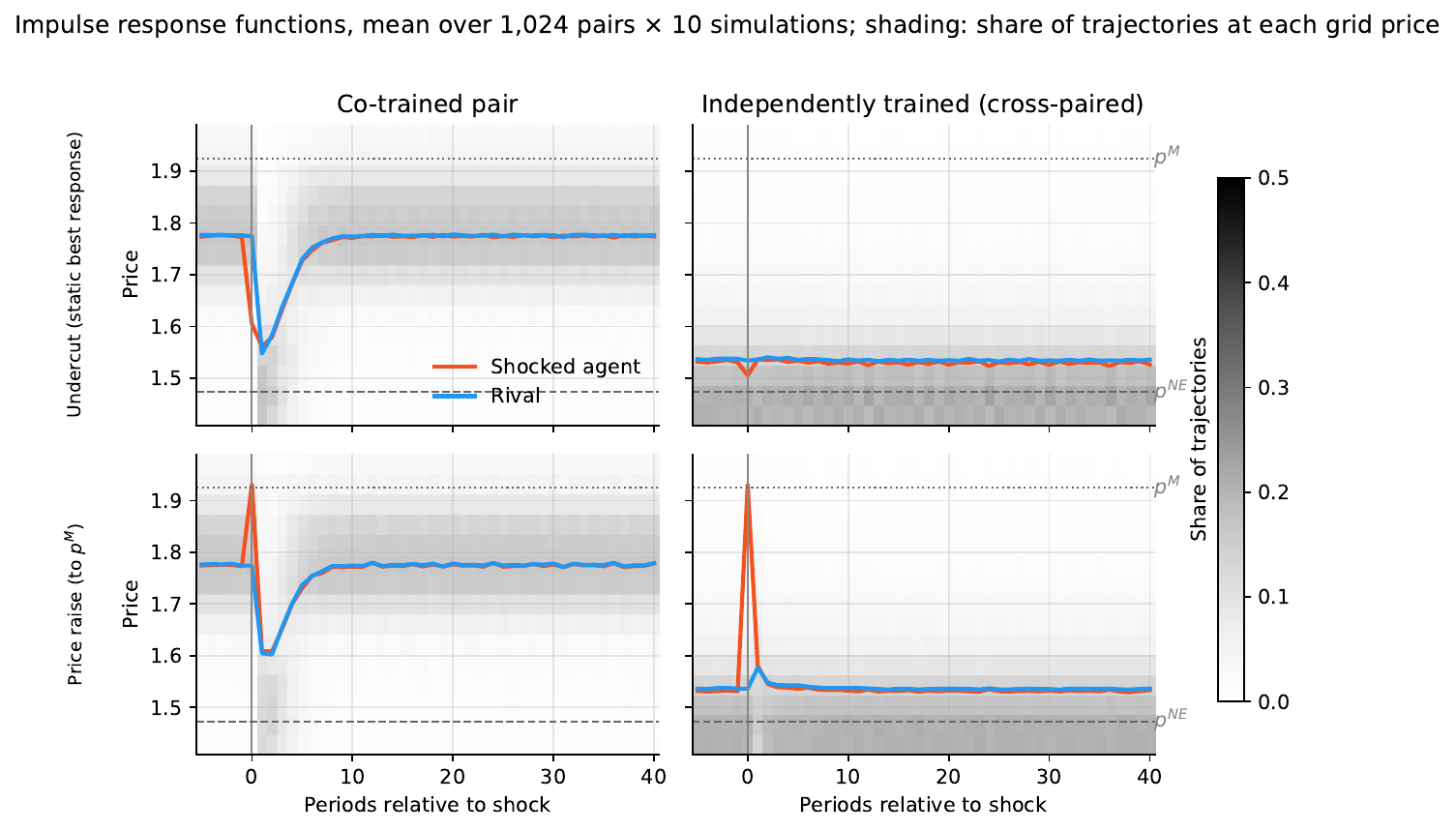}
\Description{Four charts of price over periods relative to a shock, with grey shading showing the share of 1,024 pairs times 10 simulations at each grid price in each period and lines for the mean. Top row: a one-period undercut; bottom row: a one-period price raise to the monopoly level. Left column, co-trained pairs: the mass stays concentrated, the rival's mean price drops after either shock and both means return to the pre-shock level within about ten periods. Right column, cross-paired agents: the means sit near the Nash level and barely react, and the mass splits between the lowest grid price and a diffuse band above it.}
\caption{Impulse-response functions: mean price (lines) and, in grey,
the share of trajectories at each grid price in each period (1,024 pairs
\ensuremath{\times} 10 simulations). Rows: downward shock (static best
response) and upward shock (to p\^{}M); columns: co-trained pair vs
cross-paired control. Within pairs (left), both shocks produce the
drop-and-return pattern; across pairs (right), the shock is invisible
against the ambient variation, and the mass splits between the floor of
the grid and a diffuse band above it, so the mean describes neither.}
\end{figure}

\subsection{Auditing the mechanism}\label{auditing-the-mechanism}

A punishment scheme, in the sense of Section 2, has a signature: it
responds to departures from a prescribed path, it tracks the opponent it
is meant to discipline, and it makes deviating unprofitable. This
section tests the converged policies for each element of that signature,
in increasing order of severity: the response carries no information
about \emph{what} it is reacting to (i); it is insensitive to
\emph{whom} it is playing against (ii); the off-path policy has no
best-response structure to carry a threat (iii); the celebrated return
dynamic is over-determined rather than selective (iv), the pivotal test,
because with a single global attractor the exhibit would follow any
perturbation; and none of this is a representational limit, because
hand-built punishment rules fit the same matrix exactly (v).

\paragraph{(i) The response carries no directional
information.}\label{i-the-response-carries-no-directional-information.}

This test reproduces, in our population, the finding of
\citet{epivent2024}, who show in the same environment that price hikes
are followed by the same price wars as price cuts. A trigger rule may
legitimately react to any departure from the prescribed path, upward
included, so a response to a cooperative shock does not by itself refute
path enforcement. The information is in the \emph{form} of the response.
When the forced shock goes \emph{up} (the shocked agent raises its price
to p\^{}M), the rival responds with the same qualitative pattern as
after an undercut: a drop below baseline (to \ensuremath{\approx} 1.51
against a base of 1.78, in 960 of 1,024 pairs) followed by the same
return (Figure 3, bottom-left). Nothing in the response encodes the
direction, the payoff relevance, or the identity of the perturbation. A
response that treats a move toward the cartel price exactly like a
defection is consistent with basin relaxation; it is not the signature
of a rule whose function is to deter defection. Epivent and Lambin note
that such a rule is admissible in equilibrium but has no counterpart in
documented human collusion, and that on a grid shifted below Nash the
same agents punish upward moves toward Nash; what the remaining audits
add is that it is also not what the learned policy carries.

\paragraph{(ii) There is no working model of the
opponent.}\label{ii-there-is-no-working-model-of-the-opponent.}

If the trained agents carried a contingent retaliation rule, an opponent
that actually punishes (grim trigger) should elicit different play than
an opponent that never reacts (constant Nash). Designing this comparison
requires care: started from a random state, a grim rule reads the
initial price as a defection and reverts to Nash at once; the two
opponents are then behaviorally identical, and identical play against
them proves nothing. The valid design initializes the joint state at the
cooperative price, so the punisher starts cooperative. The result is
sharper than mere blindness: \textbf{1,577 of 2,048 trained agents
(77\%) undercut the grim punisher within four periods}, triggering the
permanent reversion its rule announces, and never re-cooperate. The
agents destroy a cooperative arrangement that an opponent stands ready
to enforce (Figure 4). In the broader nine-strategy battery (2,048
agents against each of nine scripted opponents), mean CI against fixed
opponents falls to 0.155 (s.d. 0.173), against 0.67 for the same
Q-tables with their training partners, and no opponent, punishing or
cooperative, recovers it: the per-opponent means run from 0.12 against
constant Nash to 0.22 against constant monopoly.

\begin{figure}
\centering
\includegraphics[keepaspectratio]{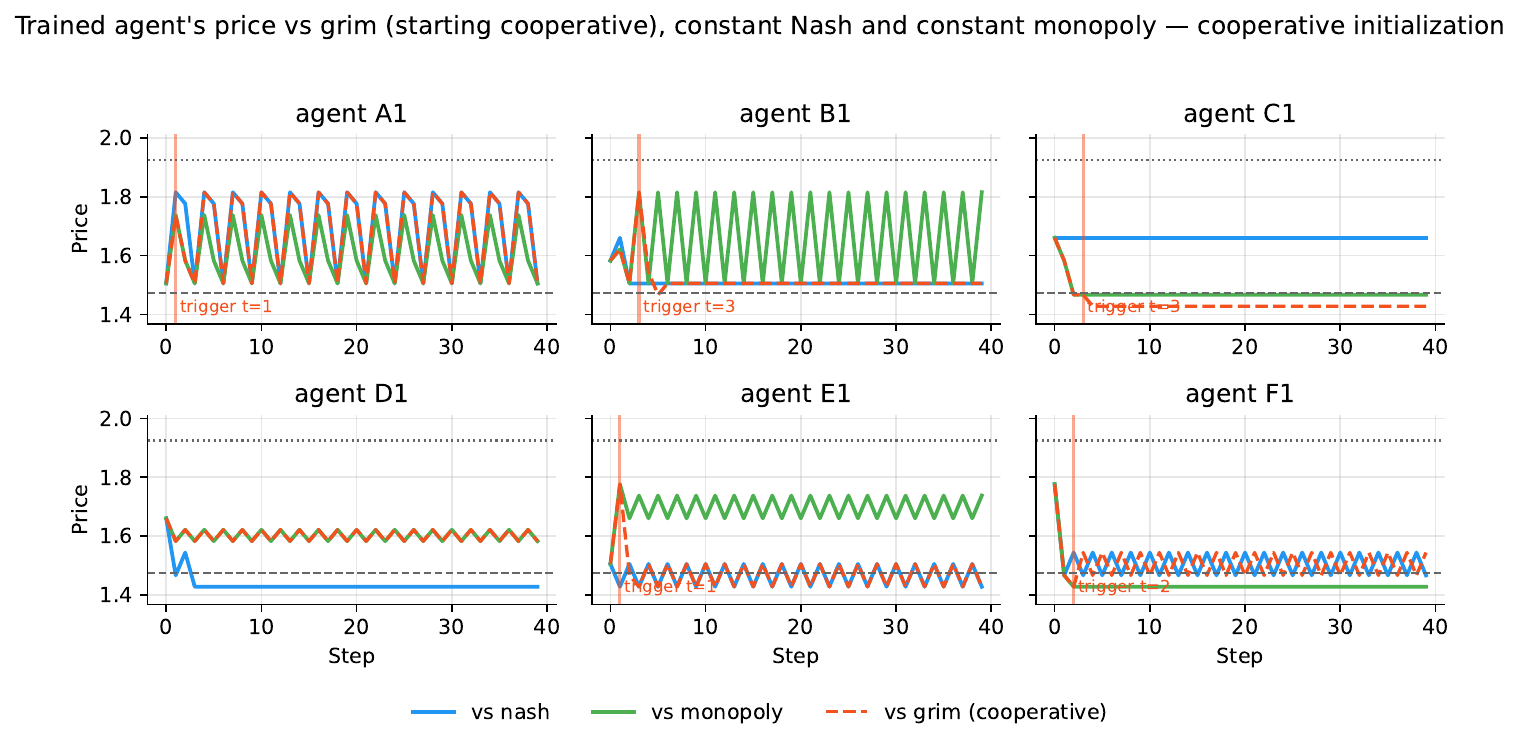}
\Description{Six small line charts, one per trained agent, of the agent's price over 40 steps against three fixed opponents: constant Nash, constant monopoly, and a cooperative grim trigger. Against the grim trigger every agent undercuts within the first few steps, marked by a vertical trigger line, and then stays at low prices; against the other opponents prices oscillate or settle at intermediate levels.}
\caption{Trained agent's price path against a grim-trigger punisher
started cooperative (dashed red), a constant-Nash pricer (blue) and a
constant-monopoly pricer (green), from a cooperative initial state; six
of the 2,048 agents shown, one panel per independent training pair
(letters A--F), first agent of each pair. Vertical lines mark the period
in which the agent's own undercut triggers grim's permanent reversion
(1,577 of 2,048 agents, t \ensuremath{\leq} 4).}
\end{figure}

\paragraph{(iii) There is no best-response logic
off-path.}\label{iii-there-is-no-best-response-logic-off-path.}

The learned policy is a 225-entry lookup table we can inspect directly.
Slicing each agent's policy at its on-path price yields its
\emph{reaction function}: price at t as a function of the rival's price
at t\ensuremath{-}1. Against the clean, monotone static best-response
curve, the 2,048 learned reaction functions are only weakly and
inconsistently related: mean Spearman \ensuremath{\rho} = \textbf{0.18},
with a range from \ensuremath{-}0.70 to 0.84 that spans every sign. The
full policy heatmaps are salt-and-pepper over the entire state space;
the only economically interpretable point is the on-path attractor
itself.

\paragraph{(iv) The return is over-determined, not
selective.}\label{iv-the-return-is-over-determined-not-selective.}

Here is the pivotal step. Under this state representation, the state
\emph{is} the previous joint action; two fixed greedy policies therefore
define a deterministic map on 225 states, and by the pigeonhole
principle every orbit is eventually periodic. Reaching \emph{some} cycle
is guaranteed by the object class; the question is whether a rule
selects the destination or the map leaves nowhere else to go.
Enumerating the full state space settles it: \textbf{914 of 1,024
trained pairs (89.3\%) have exactly one attractor whose basin is all 225
states} (mean dominant-basin share 99.3\%, mean 1.13 attractors per
pair), and the exact-return rate after a forced deviation is
\textbf{99.2\% (5,079 of 5,120 trials)}. This is the structural reason
for (i): with one globally attracting cycle, the orbit after a defection
and the orbit after a cooperative shock cannot differ. A random-policy
null locates the behavior: random Q-tables drawn on the same value range
fragment the space (1,024 random pairs: 3.39 attractors, dominant basin
77.5\%, Mann--Whitney p = 5 \ensuremath{\times} 10\ensuremath{^{-286}})
and return only \textbf{70.4\%} of the time (3,605 of 5,120 trials,
Fisher p \textless{} 10\ensuremath{^{-300}}), so the global basin is not
a property of the state space alone. What training adds is \textbf{speed
and placement}: 5.14 steps to return against 14.08 (p \textless{}
10\ensuremath{^{-300}}), and 94.2\% of returns land at CI \textgreater{}
0.3 against 68.2\% for random pairs (Figure 5). Speed and placement
without selectivity are calibration, not deterrence: the trained pair
returns to its path from anywhere and for any reason, which is what a
basin does and not what a trigger rule does.

\begin{figure}
\centering
\includegraphics[keepaspectratio]{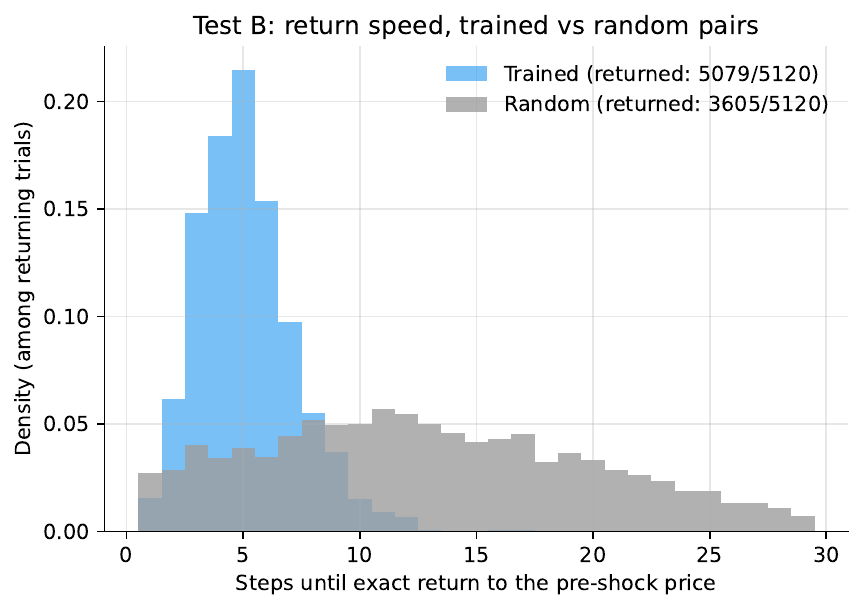}
\Description{Histogram of the number of steps until exact return to the pre-shock price, trained pairs versus random Q-tables. Trained pairs concentrate between 2 and 8 steps with 5,079 of 5,120 trials returning; random pairs spread out to 30 steps with 3,605 of 5,120 returning.}
\caption{Null model, return speed: steps until exact return to the
pre-shock price, among returning trials. Trained pairs (blue) return in
5.14 steps on average and in 5,079 of 5,120 trials; random Q-table pairs
(gray, 1,024 pairs \ensuremath{\times} 5 initializations) take 14.08
steps and return in 3,605 of 5,120 (70.4\%). Fisher p \textless{}
10\ensuremath{^{-300}} on the rates, Mann--Whitney p \textless{}
10\ensuremath{^{-300}} on the speed; the basin shares (99.3\% vs 77.5\%)
are reported in the text.}
\end{figure}

\paragraph{(v) The absence is not
architectural.}\label{v-the-absence-is-not-architectural.}

One might object that a 225-state lookup table cannot express
punishment, making its absence trivial. We check this by construction.
We hand-build two policies in the identical 225 \ensuremath{\times} 15
matrix, with the same state indexing as the trained agents: a reactive
rule (play Nash whenever the rival's last price was below a threshold,
otherwise cooperate, so punishment lasts as long as the deviation and is
then forgiven) and a grim rule (the same trigger, but once the agent's
own last price is Nash it stays at Nash, so its own past price serves as
a phase flag and the punishment is permanent). Simulated in the same
environment against a scripted rival that cooperates, undercuts to the
grid minimum for five periods and returns, from a cooperative initial
state, the reactive policy punishes for exactly the five periods of the
deviation and returns with the rival, and the grim policy punishes and
never returns (Figure 6). Their policy heatmaps show the threshold
structure as clean bands, in direct contrast with the salt-and-pepper
learned policies. The representational capacity is there; the learning
process did not use it.

\begin{figure}
\centering
\includegraphics[keepaspectratio]{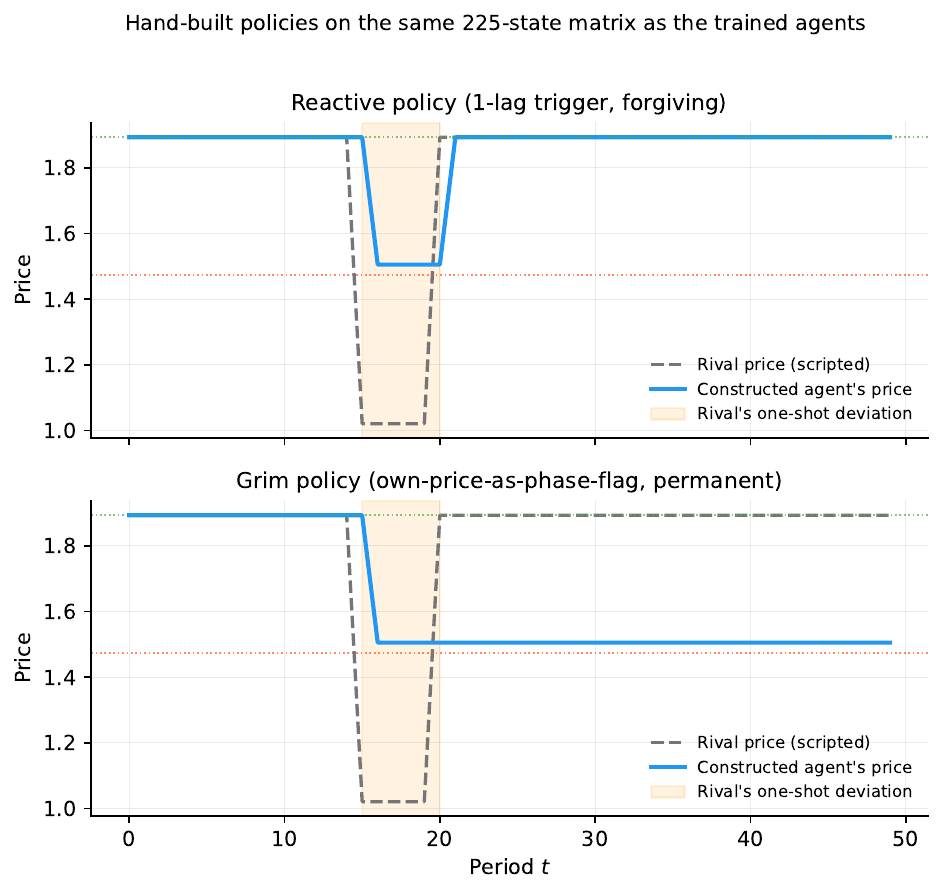}
\Description{Two line charts of price over 50 periods for hand-built policies facing a scripted rival that deviates once for four periods. Top: a reactive one-lag policy cuts its price during the deviation and returns to the high price afterwards. Bottom: a grim policy cuts its price at the deviation and stays low permanently while the rival returns to the high price.}
\caption{Capacity control: hand-built policies in the same 225
\ensuremath{\times} 15 matrix as the trained agents, simulated in the
same environment against a scripted rival that undercuts for five
periods (shaded) from a cooperative joint state. Top: the reactive rule
punishes for the duration of the deviation and forgives. Bottom: the
grim rule, using its own past price as a phase flag, punishes
permanently. Dotted lines mark the Nash and cooperative prices.}
\end{figure}

\subsection{The dyad is the unit: nothing survives outside the
pair}\label{the-dyad-is-the-unit-nothing-survives-outside-the-pair}

If what training produces is a pair-specific basin rather than
individual strategies, coordination should not survive rematching, just
as, in partner-to-stranger comparisons in the laboratory, behavior that
is a personal norm travels and behavior that is a pair-specific
convention does not. \citet[Section VI.B]{calvano2020} ran this rematch
with learning on and saw profit gains fall from 85\% to about 20\%
before re-learning partially restored them; with learning off, the
collapse is the whole story. Agents trained under identical protocols
but in different runs, when cross-matched and evaluated greedily with
the long-run price averaged over all 225 initial states, collapse from
within-pair \textbf{CI = 0.669 to CI = 0.134} across 2,000 random
matchings, an 80\% loss, with \textbf{7.4\% of matchings pricing below
Nash} (Figure 7); and matchings between pairs that had converged to the
\emph{same} price collapse just as far (0.138), so what fails to
transfer is not the price level but the off-path structure that would
return to it. The cross-paired impulse response makes the point
visually: there is no plateau to defend, prices hover noisily around
p\^{}NE, and the forced deviation is invisible against the ambient
variation (Figure 3, right column). In machine-learning terms, this is
an out-of-distribution failure: each policy was fit to the distribution
of states its particular partner generates, and a new partner, even one
trained under an identical protocol, moves the test distribution. The
pair has overfit to itself, the diagnosis of \citet{eschenbaum2022},
here at scale on Calvano et al.'s 15-price grid. Collusion in this
environment is not a strategy an agent \emph{has}; it is a convention a
pair \emph{keeps}.

\begin{figure}
\centering
\includegraphics[keepaspectratio]{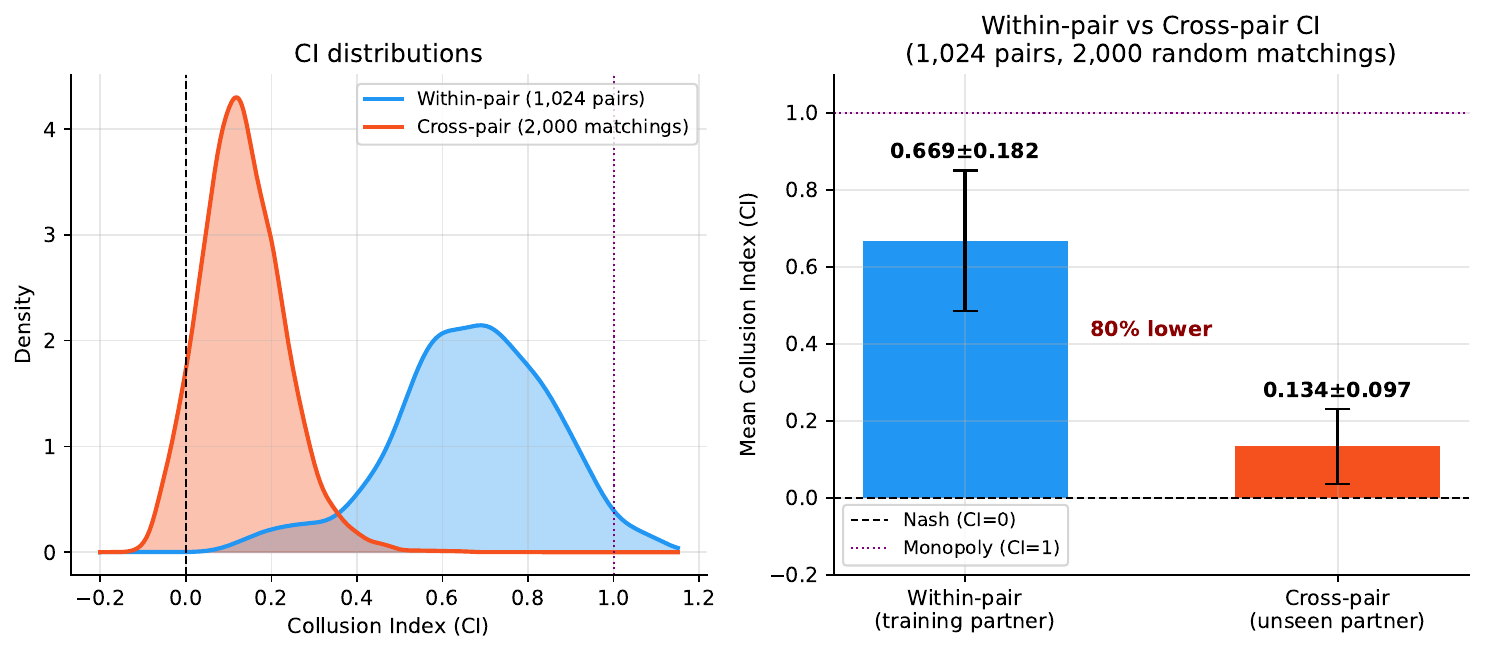}
\Description{Left: density curves of the collusion index for 1,024 within-pair evaluations, centred near 0.67, and for 2,000 cross-pair matchings, centred near 0.13. Right: bar chart of the two means with standard deviation bars, 0.669 within-pair versus 0.134 cross-pair, labelled 80 percent lower, with reference lines at Nash and monopoly.}
\caption{Within-pair vs cross-pair collusion index: distributions (left)
and means \ensuremath{\pm} 1 s.d. (right). CI collapses from 0.669 to
0.134 (\ensuremath{-}80\%) when independently trained agents are
cross-matched (1,024 pairs, 2,000 random matchings).}
\end{figure}

\subsection{What is deterred, and what is
not}\label{what-is-deterred-and-what-is-not}

\paragraph{Deterrence-in-effect holds for the deviation Calvano et
al.~test.}\label{deterrence-in-effect-holds-for-the-deviation-calvano-et-al.-test.}

We first run their own profitability calculation on our pairs: given
what the rival actually does after the forced one-period undercut, does
the deviation pay? It does not, in \textbf{946 of 1,024 pairs (92\%)}:
the one-period gain (mean +0.035) is outweighed by the discounted losses
of the post-deviation phase (mean \ensuremath{-}0.208), for a mean net
of \textbf{\ensuremath{-}0.173} (\ensuremath{\delta} = 0.95). In the 78
pairs where it pays, the rival's response exists but is too mild to
offset the one-period gain. On this point we replicate Calvano et al.

\paragraph{But the deterred deviation is not the relevant
one.}\label{but-the-deterred-deviation-is-not-the-relevant-one.}

The positive CE-gap of Section 4.1 can be made concrete. Fixing each
partner's actual greedy policy and solving the induced 225-state MDP
exactly (\ensuremath{\delta} = 0.95), the optimal best response, against
a partner who \emph{reacts}, improves on converged discounted profit by
\textbf{17.5\% on average} (median 16.2\%, maximum 93.7\%; 124 of the
2,048 agents are at an exact best response, and in only one pair are
both) (Figure 8). This is a demanding test (no learning agent computes
exact DP), and we use it not to demand optimality but to size the wedge:
the profile leaves, on average, a sixth of the achievable payoff on the
table against the very partner it converged with. The juxtaposition
resolves the apparent tension: what the converged dyad deters is the
\emph{naive} deviation, a single-period undercut followed by a return to
the old policy, which is precisely the deviation no optimizer would
choose; the \emph{optimal} deviation policy, which anticipates the
partner's reaction, profits by 17.5\%. Neither agent ever samples the
deviations that would profit it. The supracompetitive prices are not
\emph{enforced}; they are merely \emph{undisturbed}: collusion survives
because deviation is never tried, not because it would not pay. This
also locates our measure relative to the \emph{Q-loss} of \citet[Section
III.D]{calvano2020}, who evaluate each limit strategy against the
rival's and report about half of their sessions in on-path Nash
equilibrium, with a mean on-path Q-loss of 0.2\%. As we read their
procedure, the Q-loss is the forfeited payoff of a one-step improvement
under the agent's own continuation, checked at on-path states; our wedge
is the value of the full optimal policy from the converged state, which
may leave the path and keep deviating. The two are consistent: no
profitable one-step deviation on path, a profitable plan off it, which
is the sense in which the path is undisturbed rather than enforced. In
the vocabulary of Calvano IJIO and of the survey by \citet{tang2024},
the deterrence is \emph{incomplete}: genuine with respect to the
one-period deviation they test, and spurious with respect to the optimal
one, in Calvano IJIO's own sense of prices sustained by not recognizing
the gain from deviating. We therefore do not claim that retaliation
plays no stabilizing role against naive deviations; we claim that the
role it plays does not extend to the deviation an optimizer would
choose, which is the one that decides whether the path is enforced.

\begin{figure}
\centering
\includegraphics[keepaspectratio]{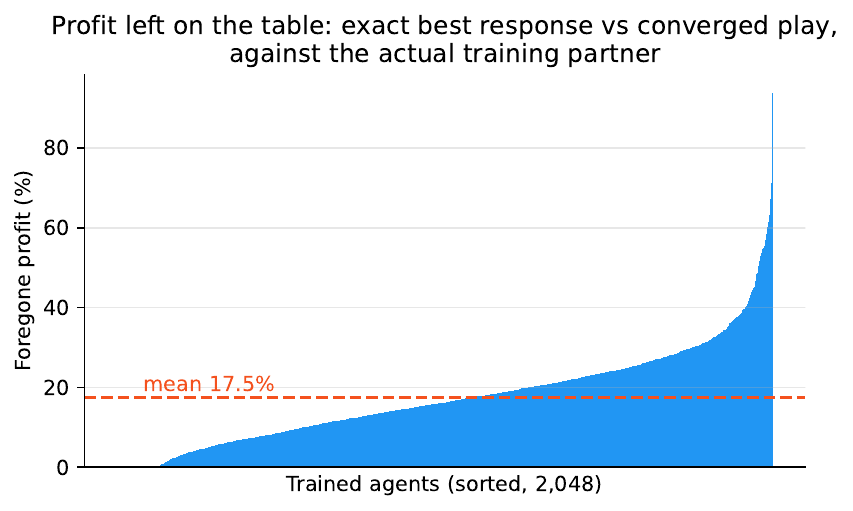}
\Description{Bar chart of foregone profit in percent for 2,048 trained agents sorted in increasing order, from zero to above 80 percent, with a dashed horizontal line at the mean of 17.5 percent.}
\caption{Exact best-response gain over converged play, per agent (2,048
agents, sorted): mean 17.5\%, median 16.2\%, maximum 93.7\%; 124 of
2,048 agents are at an exact best response.}
\end{figure}

\subsection{Why: the visitation funnel}\label{why-the-visitation-funnel}

One mechanism explains both halves of the results (why the audit finds
no scheme, and why nothing survives outside the pair), and it is visible
in training. As \ensuremath{\varepsilon} decays, state--action
visitation concentrates on a narrow neighborhood of the joint
trajectory; off-path entries of the Q-tables retain their initialization
values or reflect a handful of samples. The policy outside the basin is,
in the operative sense, untrained, which is why it behaves like the
random-map null model wherever the basin's local calibration runs out.

The funnel makes a comparative-statics prediction: enlarging the state
space, by refining the price grid, spreads the same exploration budget
over more off-path states and should weaken coordination. The prediction
holds. Holding the training budget at 2M steps and the grid span fixed,
and varying the grid from 15 to 200 points (1,024 pairs per level), mean
CI declines monotonically from 0.64 at Calvano's 15 to 0.33 (50), 0.24
(100) and 0.23 (200 points), consistent with the granularity sensitivity
in \citet{klein2021}, and prices remain supracompetitive even at the
finest grid (Figure 9). The dispersion collapses with the mean: the
standard deviation across pairs falls from 0.13 at 15 points to below
0.003 at 100 and 200, so that on fine grids every pair lands in the same
place. The basin account thus predicts evidence beyond the audit it was
built to explain. And the same exploration decay that stabilizes the
collusive path removes the experience that would test it: the funnel
builds the basin and then seals it.

\begin{figure}
\centering
\includegraphics[keepaspectratio]{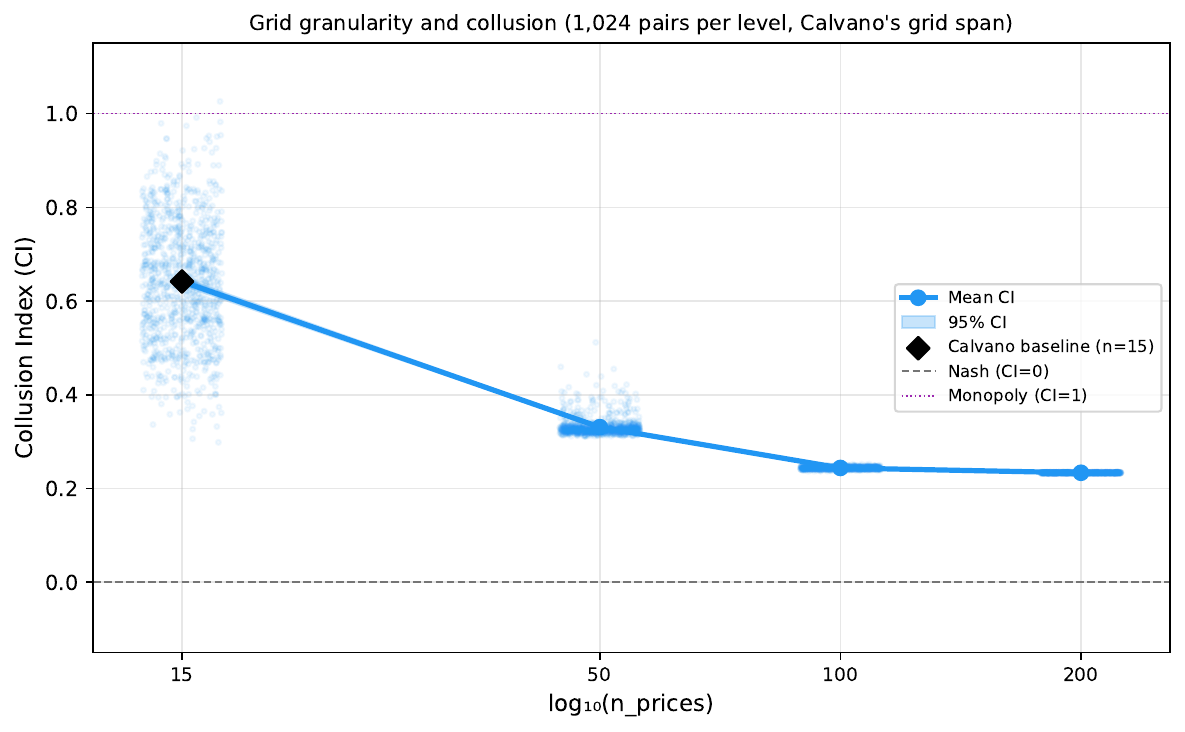}
\Description{Line chart of mean collusion index against the number of prices on a logarithmic axis, 15, 50, 100 and 200, with the individual pairs shown as a cloud of points per level. The mean falls from about 0.64 at 15 prices to about 0.33 at 50 and about 0.23 at 100 and 200, with reference lines at Nash and monopoly.}
\caption{Grid granularity: mean CI (1,024 pairs per level, 2M steps,
Calvano's grid span) declines monotonically as the price grid is refined
from 15 to 200 points, and the dispersion across pairs collapses with
it; the black diamond marks Calvano's 15-point baseline, and the 95\%
band is narrower than the line.}
\end{figure}

\section{Discussion}\label{discussion}

\paragraph{What the evidence
supports.}\label{what-the-evidence-supports.}

The converged objects in this environment are not individual pricing
strategies sustained by contingent retaliation. They are two halves of a
jointly carved attractor: fast and stable at home, structureless
everywhere else. None of the signatures of a punishment scheme we test
for is present, while the capacity control confirms the architecture
could have expressed them. The natural theoretical name for this object
is a \emph{self-confirming equilibrium} \citep{fudenberg1993}: each
agent's beliefs are correct on the realized path and never tested off
it, so play is optimal against beliefs that no experience contradicts,
without being optimal against the opponent's actual strategy. We offer
this as an interpretation, not a theorem. The dynamics that produce such
an object have a candidate description in \citet{banchio2022}, in a
stateless setting: the value estimates of co-learning agents are
correlated through the joint action, and the correlation, not any
threat, holds the profile together; the self-confirming equilibrium is
what that coupling leaves behind once exploration stops. Our capacity
control speaks to their conjecture that memory, by enabling
reward--punishment strategies, would substantially increase collusion:
with one period of memory the capacity is present and unused. We also
follow \citet{denboer2022} in preferring the language of optimality to
that of threats: algorithms are neither tempted nor deterred, and asking
whether a converged policy is a best response to its partner, as Section
4.5 does, is the question that remains well posed once that vocabulary
is set aside. Put in Lucas-critique terms: the learned rule is
calibrated to one particular environment, and here the partner \emph{is}
the environment; change the partner and the rule loses its object.

\paragraph{\texorpdfstring{Reconciling with
\citet{calvano2023}.}{Reconciling with @calvano2023.}}\label{reconciling-with-calvano2023.}

Calvano IJIO classify their converged profiles as genuinely collusive on
impulse-response evidence, and the 2020 paper verifies approximate
incentive compatibility against a single deviation; our exact-DP audit
finds mean best-response gains of 17.5\%, even though the
single-deviation profitability test of the 2020 paper \emph{replicates}
on our pairs (Section 4.5). To rule out the training regime as the
explanation, we retrained five pairs to Calvano's own stopping rule (no
episode resets, single horizon, policy stable before termination) and
reran the entire audit chain. Every result survives: the
impulse-response pattern reproduces in 5 of 5 pairs, the cooperative
grim punisher is undercut by 4 of 5 agents (the fifth never triggers
it), the one-period deviation remains unprofitable in 5 of 5, and the
exact best response still gains \textbf{10.4\% on average}. Training to
convergence \emph{narrows} the incentive wedge, from 17.5\% to 10.4\%,
but does not close it. The remaining divergence from the ``genuine
collusion'' classification lies primarily in the deviation class checked
(one-period undercut versus exact dynamic best response), not in the
training regime.

\paragraph{Implications for the literature and for
policy.}\label{implications-for-the-literature-and-for-policy.}

The impulse-response figure has served as the literature's strongest
exhibit for treating learned supracompetitive pricing as collusion in
the legal-economic sense. Our results do not overturn the figure; they
reinterpret it. A drop-and-return after any perturbation is what a
policy pair with a single global attractor does, and studies inferring
mechanisms from on-path behavior alone risk attributing to strategy what
is, mechanically, basin geometry. The consumer-welfare concern is not
dissolved: prices are robustly supracompetitive (CI \ensuremath{\approx}
0.67 within pairs). What changes is the remedial reading. The
``strategy'' does not transfer to any opponent absent from training, and
the absence of a contingent scheme cuts \emph{against} mapping the dyad
onto agreement-like legal categories. Our evidence licenses only a
frozen-policy claim: disturbing the pairing (entry, rotation,
retraining) destroys coordination among the converged policies.
\citet[Section VI.B]{calvano2020} show that continued learning re-carves
a basin after recomposition, over tens of thousands of further periods;
what our data add is that nothing is carried into that process by the
policies themselves. The enforcement problem is not smaller; it is
relocated, from detecting threats that are not there to disturbing
pairings that re-form only at the cost of re-learning at live prices.

\paragraph{Implications for machine-learning pricing at
large.}\label{implications-for-machine-learning-pricing-at-large.}

Pricing is increasingly delegated to learning systems, and algorithm
adoption measurably changes price competition
\citep{brown2023, assad2024}. For that setting our results carry a
warning and a toolkit. The warning: any claim that deployed pricing
agents have learned to coordinate, or to punish, inherits the evaluation
problem documented here, because market data record only the on-path
behavior of one grouping of algorithms, and on-path behavior alone
cannot identify a mechanism. A pricing system that performs well against
the agents it was trained with has demonstrated a property of the
\emph{ensemble}, not of the policy; our cross-pairing result measures
how large that gap can be. The toolkit: every protocol used here
(stranger re-matching, directional shocks, rule-based opponents,
random-policy null models, exploitability accounting) is defined on
frozen policies and portable to any multi-agent pricing stack, including
deep RL systems, as a pre-deployment standard for a question on-path
monitoring cannot answer: whether apparent coordination is a strategy
the agent carries, or a property of the ensemble that trained together.

\paragraph{Limitations.}\label{limitations.}

The core population is 1,024 co-trained pairs on Calvano's grid. The
main population trains a fixed 2M steps with episode resets rather than
to Calvano's stability criterion; the five-pair rerun under the faithful
regime replicates the entire audit chain, with the wedge narrowing to
10.4\%, and larger regime-faithful populations remain desirable. Our
claims are about tabular Q-learning in the Calvano environment at its
baseline parameterization (one demand specification, one 15-point grid,
\ensuremath{\delta} = 0.95). The protocol-level audits (up-shock,
opponent battery, cross-pairing) are portable to function-approximation
agents; the eventual-periodicity argument behind the null model is not,
and whether deep agents learn genuinely contingent responses is a
question our protocol is designed to answer, not presuppose. Within that
scope, none of the signatures of deterrence we test for is present, and
what is present is basin geometry.

\section{Conclusion}\label{conclusion}

We replicated the central exhibit of the algorithmic collusion
literature and asked the question the exhibit was taken to answer: is
there a punishment mechanism? None of five independent audits finds one.
The response to deviation carries no directional information; faced with
a punisher whose cooperation was there to be kept, the agents defect and
absorb the punishment; their reaction functions are only weakly related
to best-response logic; their celebrated return-to-baseline is
over-determined, since nine pairs in ten have a single attractor
covering the whole state space; and none of this is explained by a
representational limit. What training produces instead is a
dyad-specific basin of attraction: re-match frozen agents from
independent runs and coordination collapses by 80\%, from CI = 0.669 to
0.134. And the path the basin holds is not incentive-compatible against
the optimal deviation, which, computed against each partner's frozen
strategy, profits by 17.5\% on average. The scheme deters only the
deviation no optimizer would choose. Collusion here is not a strategy
either agent has; it is a convention the pair keeps. This reading
unifies what the literature had measured separately: the
pair-specificity Calvano et al.~observed on rematching, the
direction-blind response Epivent and Lambin documented, the gap between
impulse patterns and equilibrium den Boer et al.~established, and the
punishment-free coupling Banchio and Mantegazza modeled.

The methodological point outlasts the specific exhibit. As pricing
migrates to learning systems, behavioral patterns produced by black-box
agents will increasingly reach regulators and courts as evidence of
algorithmic coordination, and such patterns invite mechanistic
over-reading: the drop-and-return \emph{looks} like retaliation, and
controls are what tell the difference. Mechanism claims about learned
behavior should be held to the standard applied here: survive an
explicit null, survive removal from the training context, survive an
incentive accounting. What the standard buys is precision about the
policy problem. The supracompetitive prices, and the harm they imply,
are robust; what is absent is the deterrence behind them, and harm and
mechanism call for different instruments: an outcome can be monitored,
but only an agent that carries a rule can be deterred. Prices in this
environment are high not because deviation is punished, but because,
once exploration has decayed, deviation is never tried, and the beliefs
that would motivate it are never corrected.

\begin{acks}
The authors thank the Funda\c{c}\~{a}o Instituto de Pesquisas Econ\^{o}micas (FIPE, Institute of Economic Research Foundation) for its financial support.
\end{acks}

\bibliographystyle{ACM-Reference-Format}
\bibliography{refs}


\begin{thebibliography}{23}


\ifx \showCODEN    \undefined \def \showCODEN     #1{\unskip}     \fi
\ifx \showISBNx    \undefined \def \showISBNx     #1{\unskip}     \fi
\ifx \showISBNxiii \undefined \def \showISBNxiii  #1{\unskip}     \fi
\ifx \showISSN     \undefined \def \showISSN      #1{\unskip}     \fi
\ifx \showLCCN     \undefined \def \showLCCN      #1{\unskip}     \fi
\ifx \shownote     \undefined \def \shownote      #1{#1}          \fi
\ifx \showarticletitle \undefined \def \showarticletitle #1{#1}   \fi
\ifx \showURL      \undefined \def \showURL       {\relax}        \fi
\providecommand\bibfield[2]{#2}
\providecommand\bibinfo[2]{#2}
\providecommand\natexlab[1]{#1}
\providecommand\showeprint[2][]{arXiv:#2}
\makeatletter
\@ifundefined{NAT@parse@date}{}{\let\NAT@parse@date@orig\NAT@parse@date}
\@ifundefined{NAT@parse@date}{}{\def\NAT@parse@date#1#2#3#4#5#6@@{\NAT@parse@date@orig#1#2#3#4#5#6@@\def\NAT@tempyear{0000}\def\NAT@tempexlab{{?}}\ifx\NAT@year\NAT@tempyear\ifx\NAT@exlab\NAT@tempexlab\def\NAT@date{[n.\,d.]}\else\edef\NAT@date{[n.\,d.]\NAT@exlab}\fi\fi}}
\makeatother

\bibitem[Abada and Lambin(2023)]%
        {abada2023}
\bibfield{author}{\bibinfo{person}{Ibrahim Abada} {and} \bibinfo{person}{Xavier
  Lambin}.} \bibinfo{year}{2023}\natexlab{}.
\newblock \showarticletitle{Artificial Intelligence: Can Seemingly Collusive
  Outcomes Be Avoided?}
\newblock \bibinfo{journal}{\emph{Management Science}} \bibinfo{volume}{69},
  \bibinfo{number}{9} (\bibinfo{year}{2023}), \bibinfo{pages}{5042--5065}.
\newblock
\href{https://doi.org/10.1287/mnsc.2022.4623}{doi:\nolinkurl{10.1287/mnsc.2022.4623}}


\bibitem[Abreu(1988)]%
        {abreu1988}
\bibfield{author}{\bibinfo{person}{Dilip Abreu}.}
  \bibinfo{year}{1988}\natexlab{}.
\newblock \showarticletitle{On the Theory of Infinitely Repeated Games with
  Discounting}.
\newblock \bibinfo{journal}{\emph{Econometrica}} \bibinfo{volume}{56},
  \bibinfo{number}{2} (\bibinfo{year}{1988}), \bibinfo{pages}{383--396}.
\newblock
\href{https://doi.org/10.2307/1911077}{doi:\nolinkurl{10.2307/1911077}}


\bibitem[Andreoni(1988)]%
        {andreoni1988}
\bibfield{author}{\bibinfo{person}{James Andreoni}.}
  \bibinfo{year}{1988}\natexlab{}.
\newblock \showarticletitle{Why Free Ride? {Strategies} and Learning in Public
  Goods Experiments}.
\newblock \bibinfo{journal}{\emph{Journal of Public Economics}}
  \bibinfo{volume}{37}, \bibinfo{number}{3} (\bibinfo{year}{1988}),
  \bibinfo{pages}{291--304}.
\newblock
\href{https://doi.org/10.1016/0047-2727(88)90043-6}{doi:\nolinkurl{10.1016/0047-2727(88)90043-6}}


\bibitem[Asker et~al\mbox{.}(2024)]%
        {asker2024}
\bibfield{author}{\bibinfo{person}{John Asker}, \bibinfo{person}{Chaim
  Fershtman}, {and} \bibinfo{person}{Ariel Pakes}.}
  \bibinfo{year}{2024}\natexlab{}.
\newblock \showarticletitle{The Impact of Artificial Intelligence Design on
  Pricing}.
\newblock \bibinfo{journal}{\emph{Journal of Economics \& Management Strategy}}
  \bibinfo{volume}{33}, \bibinfo{number}{2} (\bibinfo{year}{2024}),
  \bibinfo{pages}{276--304}.
\newblock
\href{https://doi.org/10.1111/jems.12516}{doi:\nolinkurl{10.1111/jems.12516}}


\bibitem[Assad et~al\mbox{.}(2024)]%
        {assad2024}
\bibfield{author}{\bibinfo{person}{Stephanie Assad}, \bibinfo{person}{Robert
  Clark}, \bibinfo{person}{Daniel Ershov}, {and} \bibinfo{person}{Lei Xu}.}
  \bibinfo{year}{2024}\natexlab{}.
\newblock \showarticletitle{Algorithmic Pricing and Competition: Empirical
  Evidence from the {German} Retail Gasoline Market}.
\newblock \bibinfo{journal}{\emph{Journal of Political Economy}}
  \bibinfo{volume}{132}, \bibinfo{number}{3} (\bibinfo{year}{2024}),
  \bibinfo{pages}{723--771}.
\newblock
\href{https://doi.org/10.1086/726906}{doi:\nolinkurl{10.1086/726906}}


\bibitem[Banchio and Mantegazza(2023)]%
        {banchio2022}
\bibfield{author}{\bibinfo{person}{Martino Banchio} {and}
  \bibinfo{person}{Giacomo Mantegazza}.} \bibinfo{year}{2023}\natexlab{}.
\newblock \bibinfo{title}{Artificial Intelligence and Spontaneous Collusion}.
\newblock \bibinfo{howpublished}{arXiv:2202.05946v5 [econ.TH]}.
\newblock
\showeprint[arxiv]{2202.05946}~[econ.TH]
\newblock
\shownote{Version 5, September 19, 2023}.


\bibitem[Brown and MacKay(2023)]%
        {brown2023}
\bibfield{author}{\bibinfo{person}{Zach~Y. Brown} {and}
  \bibinfo{person}{Alexander MacKay}.} \bibinfo{year}{2023}\natexlab{}.
\newblock \showarticletitle{Competition in Pricing Algorithms}.
\newblock \bibinfo{journal}{\emph{American Economic Journal: Microeconomics}}
  \bibinfo{volume}{15}, \bibinfo{number}{2} (\bibinfo{year}{2023}),
  \bibinfo{pages}{109--156}.
\newblock
\href{https://doi.org/10.1257/mic.20210158}{doi:\nolinkurl{10.1257/mic.20210158}}


\bibitem[Calvano et~al\mbox{.}(2020)]%
        {calvano2020}
\bibfield{author}{\bibinfo{person}{Emilio Calvano}, \bibinfo{person}{Giacomo
  Calzolari}, \bibinfo{person}{Vincenzo Denicol{\`o}}, {and}
  \bibinfo{person}{Sergio Pastorello}.} \bibinfo{year}{2020}\natexlab{}.
\newblock \showarticletitle{Artificial Intelligence, Algorithmic Pricing, and
  Collusion}.
\newblock \bibinfo{journal}{\emph{American Economic Review}}
  \bibinfo{volume}{110}, \bibinfo{number}{10} (\bibinfo{year}{2020}),
  \bibinfo{pages}{3267--3297}.
\newblock
\href{https://doi.org/10.1257/aer.20190623}{doi:\nolinkurl{10.1257/aer.20190623}}


\bibitem[Calvano et~al\mbox{.}(2023)]%
        {calvano2023}
\bibfield{author}{\bibinfo{person}{Emilio Calvano}, \bibinfo{person}{Giacomo
  Calzolari}, \bibinfo{person}{Vincenzo Denicol{\`o}}, {and}
  \bibinfo{person}{Sergio Pastorello}.} \bibinfo{year}{2023}\natexlab{}.
\newblock \showarticletitle{Algorithmic Collusion: Genuine or Spurious?}
\newblock \bibinfo{journal}{\emph{International Journal of Industrial
  Organization}}  \bibinfo{volume}{90} (\bibinfo{year}{2023}),
  \bibinfo{pages}{102973}.
\newblock
\href{https://doi.org/10.1016/j.ijindorg.2023.102973}{doi:\nolinkurl{10.1016/j.ijindorg.2023.102973}}


\bibitem[den Boer(2023)]%
        {denboer2023}
\bibfield{author}{\bibinfo{person}{Arnoud~V. den Boer}.}
  \bibinfo{year}{2023}\natexlab{}.
\newblock \bibinfo{title}{Algorithmic Collusion: A Mathematical Definition and
  Research Agenda for the {OR/MS} Community}.
\newblock \bibinfo{howpublished}{SSRN Working Paper 4636488}.
\newblock
\href{https://doi.org/10.2139/ssrn.4636488}{doi:\nolinkurl{10.2139/ssrn.4636488}}


\bibitem[den Boer et~al\mbox{.}(2026)]%
        {denboer2022}
\bibfield{author}{\bibinfo{person}{Arnoud~V. den Boer},
  \bibinfo{person}{Janusz~M. Meylahn}, {and} \bibinfo{person}{Maarten~Pieter
  Schinkel}.} \bibinfo{year}{2026}\natexlab{}.
\newblock \showarticletitle{Artificial Collusion: Examining Supracompetitive
  Pricing by {Q}-Learning Algorithms}.
\newblock \bibinfo{journal}{\emph{Management Science}} (\bibinfo{year}{2026}).
\newblock
\href{https://doi.org/10.1287/mnsc.2024.08557}{doi:\nolinkurl{10.1287/mnsc.2024.08557}}
\newblock
\shownote{Articles in Advance, published online June 9, 2026}.


\bibitem[Epivent and Lambin(2024)]%
        {epivent2024}
\bibfield{author}{\bibinfo{person}{Andr{\'e}a Epivent} {and}
  \bibinfo{person}{Xavier Lambin}.} \bibinfo{year}{2024}\natexlab{}.
\newblock \showarticletitle{On Algorithmic Collusion and Reward--Punishment
  Schemes}.
\newblock \bibinfo{journal}{\emph{Economics Letters}}  \bibinfo{volume}{237}
  (\bibinfo{year}{2024}), \bibinfo{pages}{111661}.
\newblock
\href{https://doi.org/10.1016/j.econlet.2024.111661}{doi:\nolinkurl{10.1016/j.econlet.2024.111661}}


\bibitem[Eschenbaum et~al\mbox{.}(2022)]%
        {eschenbaum2022}
\bibfield{author}{\bibinfo{person}{Nicolas Eschenbaum}, \bibinfo{person}{Filip
  Mellgren}, {and} \bibinfo{person}{Philipp Zahn}.}
  \bibinfo{year}{2022}\natexlab{}.
\newblock \bibinfo{title}{Robust Algorithmic Collusion}.
\newblock
\showeprint[arxiv]{2201.00345}~[econ.GN]


\bibitem[Friedman(1971)]%
        {friedman1971}
\bibfield{author}{\bibinfo{person}{James~W. Friedman}.}
  \bibinfo{year}{1971}\natexlab{}.
\newblock \showarticletitle{A Non-cooperative Equilibrium for Supergames}.
\newblock \bibinfo{journal}{\emph{Review of Economic Studies}}
  \bibinfo{volume}{38}, \bibinfo{number}{1} (\bibinfo{year}{1971}),
  \bibinfo{pages}{1--12}.
\newblock
\href{https://doi.org/10.2307/2296617}{doi:\nolinkurl{10.2307/2296617}}


\bibitem[Fudenberg and Levine(1993)]%
        {fudenberg1993}
\bibfield{author}{\bibinfo{person}{Drew Fudenberg} {and}
  \bibinfo{person}{David~K. Levine}.} \bibinfo{year}{1993}\natexlab{}.
\newblock \showarticletitle{Self-Confirming Equilibrium}.
\newblock \bibinfo{journal}{\emph{Econometrica}} \bibinfo{volume}{61},
  \bibinfo{number}{3} (\bibinfo{year}{1993}), \bibinfo{pages}{523--545}.
\newblock
\href{https://doi.org/10.2307/2951716}{doi:\nolinkurl{10.2307/2951716}}


\bibitem[Green and Porter(1984)]%
        {green1984}
\bibfield{author}{\bibinfo{person}{Edward~J. Green} {and}
  \bibinfo{person}{Robert~H. Porter}.} \bibinfo{year}{1984}\natexlab{}.
\newblock \showarticletitle{Noncooperative Collusion under Imperfect Price
  Information}.
\newblock \bibinfo{journal}{\emph{Econometrica}} \bibinfo{volume}{52},
  \bibinfo{number}{1} (\bibinfo{year}{1984}), \bibinfo{pages}{87--100}.
\newblock
\href{https://doi.org/10.2307/1911462}{doi:\nolinkurl{10.2307/1911462}}


\bibitem[Hansen et~al\mbox{.}(2021)]%
        {hansen2021}
\bibfield{author}{\bibinfo{person}{Karsten~T. Hansen},
  \bibinfo{person}{Kanishka Misra}, {and} \bibinfo{person}{Mallesh~M. Pai}.}
  \bibinfo{year}{2021}\natexlab{}.
\newblock \showarticletitle{Frontiers: Algorithmic Collusion: Supra-competitive
  Prices via Independent Algorithms}.
\newblock \bibinfo{journal}{\emph{Marketing Science}} \bibinfo{volume}{40},
  \bibinfo{number}{1} (\bibinfo{year}{2021}), \bibinfo{pages}{1--12}.
\newblock
\href{https://doi.org/10.1287/mksc.2020.1276}{doi:\nolinkurl{10.1287/mksc.2020.1276}}


\bibitem[Harrington(2018)]%
        {harrington2018}
\bibfield{author}{\bibinfo{person}{Joseph~E. Harrington}.}
  \bibinfo{year}{2018}\natexlab{}.
\newblock \showarticletitle{Developing Competition Law for Collusion by
  Autonomous Artificial Agents}.
\newblock \bibinfo{journal}{\emph{Journal of Competition Law \& Economics}}
  \bibinfo{volume}{14}, \bibinfo{number}{3} (\bibinfo{year}{2018}),
  \bibinfo{pages}{331--363}.
\newblock
\href{https://doi.org/10.1093/joclec/nhy016}{doi:\nolinkurl{10.1093/joclec/nhy016}}


\bibitem[Klein(2021)]%
        {klein2021}
\bibfield{author}{\bibinfo{person}{Timo Klein}.}
  \bibinfo{year}{2021}\natexlab{}.
\newblock \showarticletitle{Autonomous Algorithmic Collusion: {Q}-Learning
  under Sequential Pricing}.
\newblock \bibinfo{journal}{\emph{RAND Journal of Economics}}
  \bibinfo{volume}{52}, \bibinfo{number}{3} (\bibinfo{year}{2021}),
  \bibinfo{pages}{538--558}.
\newblock
\href{https://doi.org/10.1111/1756-2171.12383}{doi:\nolinkurl{10.1111/1756-2171.12383}}


\bibitem[Lanctot et~al\mbox{.}(2017)]%
        {lanctot2017}
\bibfield{author}{\bibinfo{person}{Marc Lanctot}, \bibinfo{person}{Vinicius
  Zambaldi}, \bibinfo{person}{Audrunas Gruslys}, \bibinfo{person}{Angeliki
  Lazaridou}, \bibinfo{person}{Karl Tuyls}, \bibinfo{person}{Julien
  P{\'e}rolat}, \bibinfo{person}{David Silver}, {and} \bibinfo{person}{Thore
  Graepel}.} \bibinfo{year}{2017}\natexlab{}.
\newblock \showarticletitle{A Unified Game-Theoretic Approach to Multiagent
  Reinforcement Learning}. In \bibinfo{booktitle}{\emph{Advances in Neural
  Information Processing Systems 30 (NIPS 2017)}}, Vol.~\bibinfo{volume}{30}.
\newblock


\bibitem[{OECD}(2017)]%
        {oecd2017}
\bibfield{author}{\bibinfo{person}{{OECD}}.} \bibinfo{year}{2017}\natexlab{}.
\newblock \bibinfo{booktitle}{\emph{Algorithms and Collusion: Competition
  Policy in the Digital Age}}.
\newblock \bibinfo{type}{{T}echnical {R}eport}. \bibinfo{institution}{OECD
  Publishing}, \bibinfo{address}{Paris}.
\newblock
\href{https://doi.org/10.1787/258dcb14-en}{doi:\nolinkurl{10.1787/258dcb14-en}}


\bibitem[Tang(2024)]%
        {tang2024}
\bibfield{author}{\bibinfo{person}{Chen Tang}.}
  \bibinfo{year}{2024}\natexlab{}.
\newblock \bibinfo{title}{Algorithmic Collusion in Dynamic Pricing: Past,
  Present, and Future}.
\newblock \bibinfo{howpublished}{SSRN Working Paper 4891632}.
\newblock
\href{https://doi.org/10.2139/ssrn.4891632}{doi:\nolinkurl{10.2139/ssrn.4891632}}


\bibitem[Watkins and Dayan(1992)]%
        {watkins1992}
\bibfield{author}{\bibinfo{person}{Christopher J. C.~H. Watkins} {and}
  \bibinfo{person}{Peter Dayan}.} \bibinfo{year}{1992}\natexlab{}.
\newblock \showarticletitle{Q-learning}.
\newblock \bibinfo{journal}{\emph{Machine Learning}} \bibinfo{volume}{8},
  \bibinfo{number}{3--4} (\bibinfo{year}{1992}), \bibinfo{pages}{279--292}.
\newblock
\href{https://doi.org/10.1007/BF00992698}{doi:\nolinkurl{10.1007/BF00992698}}


\end{thebibliography}

\end{document}